\documentclass[aps,prl,twocolumn,showpacs,groupedaddress]{revtex4}  
\usepackage{graphicx}  
\usepackage{dcolumn}   
\usepackage{bm}        
\usepackage{amssymb,amsmath}   
\usepackage{amsthm}
\usepackage{subfigure}

\begin{document}

\title{Network Quotients: Structural Skeletons of Complex Systems}

\author{Yanghua Xiao$^{1}$}
\author{Ben D. MacArthur$^{2}$}
\author{Hui Wang$^{3}$}
\author{Momiao Xiong$^{4,5}$}
\author{Wei Wang$^{1}$}

\affiliation{\vspace{0.1 in}\vspace{0.1 in}}

\affiliation{$^{1}$Department of Computing and Information Technology, Fudan University, Shanghai 200433, PR China}

\affiliation{$^{2}$Bone and Joint Research Group, Centre for Human Development, Stem Cells and Regeneration, Institute of Developmental Sciences, University of Southampton, Southampton, SO16 6YD, UK}

\affiliation{$^{3}$Business School, University of Shanghai for Science and Technology, Shanghai 200093, PR China}

\affiliation{$^{4}$Theoretical Systems Biology Lab, School of Life Science, Fudan University, Shanghai 200433, PR China}

\affiliation{$^{5}$Human Genetics Center, University of Texas Health Science Center at Houston, Houston TX 77225, USA}

\date{\today}

\begin{abstract}

A defining feature of many large empirical networks is their
intrinsic complexity. However, many networks also contain a large
degree of structural repetition. An immediate question then arises:
can we characterize essential network complexity while excluding
structural redundancy?

In this article we utilize inherent network symmetry to collapse all
redundant information from a network, resulting in a coarse-graining
which we show to carry the essential structural information of the
`parent' network. In the context of algebraic combinatorics, this
coarse-graining is known as the \emph{quotient}.  We systematically
explore the theoretical properties of network quotients and
summarize key statistics of a variety of `real-world' quotients with
respect to those of their parent networks. In particular, we find
that quotients can be substantially smaller than their parent
networks yet typically preserve various key functional properties
such as complexity (heterogeneity and hubs vertices) and
communication (diameter and mean geodesic distance), suggesting that
quotients constitute the essential structural skeleton of their
parent network. We summarize with a discussion of potential uses of
quotients in analysis of biological regulatory networks and ways in
which using quotients can reduce the computational complexity of
network algorithms.

\end{abstract}


\maketitle

\section{Introduction}
Many physical systems -- from the world-wide web to scientific
collaborations and biochemical reactions inside cells -- can be
modeled as networks. The ubiquity of empirical networks has
generated increasing interest in their study over the last decade
during which much progress has been made toward elucidating general
network organizational principles beyond the specific details of
individual
systems\cite{albertSM,newman,kleinberg,amaral,watts,barabasiBA}.
Structural properties which are commonly found in many disparate
networks include: the `small-world' property\cite{watts}; the
scale-free distribution of vertex degrees\cite{barabasiBA};
hierarchical modularity\cite{ravasz}; network construction from
motifs\cite{motif}; assortative mixing\cite{Newman2002}; and
self-similarity\cite{song} amongst others. Together,
investigation of generic structural properties such as these may be
thought of as an attempt to understand network \emph{complexity}\cite{bara_tm}.

In order to find simplicity in this complexity some authors have
attempted to extract network `skeletons': related networks which
capture essential structural features of the system from which they
are derived, but are simpler in some quantitative way. Existing
network skeletons include for instance, the fractal
skeleton\cite{fractal_skeleton}, which is responsible for fractal
scaling; and the communication skeleton\cite{tree}, which is
responsible for the majority of communication flow through the
network. Such skeletons are generally formed with respect to a given
property, for example fractality or communication, and thus do not
represent a structural skeleton in the strongest sense. In this
article we propose an alternative skeleton which formally captures
all essential structural information, and which can be significantly
smaller than the original network from which it is derived. The
method we use is based upon utilizing inherent network symmetry.

Although almost all large random networks are
asymmetric\cite{bollobasRG}, many empirical real-world networks are
surprisingly richly symmetric\cite{sym1,sym2,xiao1,xiao2}. This
symmetry commonly results from the presence of locally tree-like or
biclique-like structures\cite{sym2,xiao2} which are present in many
empirical networks, and derive naturally from elementary growth
processes such as growth with preferential attachment\cite{sym2} and
growth with similar linkage pattern\cite{xiao2}. However,
despite a rich abstract theory of graph
symmetry\cite{lauri,Biggs,lwb,Godsil}, the symmetry structure of
complex real-world networks has not yet been explored extensively.

Intuitively, a network is symmetric if two or more of its vertices
can be permuted without altering vertex adjacency. Symmetric
networks therefore necessarily contain a certain degree structural
redundancy in that they possess multiple vertices which play the
same structural role. Thus network symmetry is strongly related to
network redundancy.In this paper we use this relationship to show
how symmetry also provides a natural means to formally exclude
redundancy while still preserving essential network structure, by
factoring out structurally identical elements.

The structure of the remainder of this paper is as follows: first we
introduce essential background material concerning network
automorphism groups, and show how a network's automorphism group may
be used to produce a coarse-grained skeleton of the network called the
\emph{quotient}. We also introduce a variation on the classical
quotient which we call the \emph{s-quotient}. Then we show that
quotients can be substantially smaller than the network from which
they are derived and explore ways in which key structural properties
are inherited by the quotient and the s-quotient from the `parent'
network. In particular, we shall examine how network heterogeneity,
degree distribution and communication properties are carried from
the parent to its quotients.

\section{Background and Definitions}
\subsection{Preliminaries}
Formally, a network is a graph $G=(V,E)$ with vertex set $V$ and
edge set $E$ in which two vertices are adjacent if there is an edge
between them. An automorphism is a permutation of the vertices of
the network which preserves adjacency, and the set of automorphisms
under composition forms a group $\textrm{Aut}(G)$. The automorphism
group of a network compactly describes its symmetry structure.
Automorphism groups can be efficiently calculated with the use of an
appropriate graph isomorphism algorithm such as the \texttt{nauty}
algorithm\cite{mckay} which we use in this study. We say that a
network which possesses a nontrivial automorphism group is
\emph{symmetric}. Previous studies have highlighted the fact that
many real-world networks possess nontrivial (and often quite large)
automorphism groups\cite{sym1,sym2,xiao1,xiao2}.

The vertices of a symmetric network can be partitioned into disjoint
equivalence classes called \emph{orbits}: for every vertex $v \in
V(G)$, $v$ belongs to the orbit
\[
\Delta(v) = \{g\cdot v \in V : g \in \textrm{Aut}(G) \}.
\]
We refer to the partitioning of the network vertices into disjoint
orbits as the \emph{automorphism partition}\cite{xiao1}. Note that
since they can be permuted without altering network structure, two
vertices in the same orbit are equivalent in the strongest possible
structural sense: they play \emph{precisely} the same structural
role in the network and cannot be distinguished from one another by
any meaningful structural measure (more formally, a vertex property
which is preserved under isomorphism is known as a \emph{vertex
invariant}; vertices in the same orbit are indistinguishable by
vertex invariants\cite{gtin}). Vertices in the same orbit therefore
possess many of the same structural properties, including the same
degree, eigenvector centrality and clustering coefficient\cite{sym1}
(for more examples, see \cite{gtin}). We therefore say that vertices
in the same group orbit are \emph{structurally equivalent}. Since
many real-world empirical networks possess a non-trivial
automorphism partition they therefore carry a significant amount of
redundant information in which more than one vertex plays the same
structural role. In addition to elucidating the precise nature of
structural repetitions in a network, the automorphism partition also
provides a convenient way to factor out these structural repetitions
by `gluing together' structurally equivalent vertices to create a
coarse-graining of the network, known as the \emph{quotient}.

\subsection{Quotients}
More formally, let $\mathbf{\Delta} = \{ \Delta_1,\Delta_2 \ldots,
\Delta_s \}$ be the automorphism partition of a network $G$. A
significant property of this partition is that it is
\emph{equitable}\cite{Godsil}: the number of neighbors in $\Delta_j$
of a vertex $v \in \Delta_i$ is a constant $q_{ij}$ ($i,j =
1,2,\ldots,s$), which depends upon $i$ and $j$ but is independent of
the choice of $v \in \Delta_i$. The \emph{quotient} $Q$ of $G$ under
the action of $\textrm{Aut}(G)$ is the multi-digraph with vertex set
$\mathbf{\Delta}$ and adjacency matrix $q_{ij}$. We refer to the
network $G$ as the \textit{parent} of $Q$, and note that network
quotients may be easily calculated using the \texttt{nauty}
algorithm\cite{mckay}.

The quotient contains all the structural information of its parent
network but, by associating structurally equivalent vertices,
formally excludes all structural repetitions.Crucially, this means
that many characteristic properties of the parent network are
preserved in the quotient (any differences are due to the fact that
the quotient only carries the unique structural features of its
parent without repetitions). Consequently, while they are often very
similar, it is the properties of the quotient, and not those of the
parent network \emph{per se}, that describe core system complexity.
For this reason, the quotient may be thought of as the structural
\emph{skeleton} of its parent.

In the context of algebraic graph
theory\cite{lauri,Biggs,lwb,Godsil}, certain properties of quotients
are well-known including, for example, the fact that the eigenvalues
of the quotient are a subset of those of its parent\cite{Godsil}.
However, previous studies of graph quotients have been largely
mathematical in nature, and have tended to focus on properties of
quotients of completely regular graphs. An investigation of the
properties of quotients of real-world networks -- which typically
contain both regular and random elements -- has not as yet been
undertaken.

\subsection{S-Quotients}
As we noted above, quotients are generally multi-digraphs (that is,
their edges are weighted and directed). This is the case even when
the parent network is simple (that is, the edges are not weighted or
directed).

\begin{figure*}
\centering
 \subfigure[]{ \label{fig:net}
\includegraphics[scale=0.15]{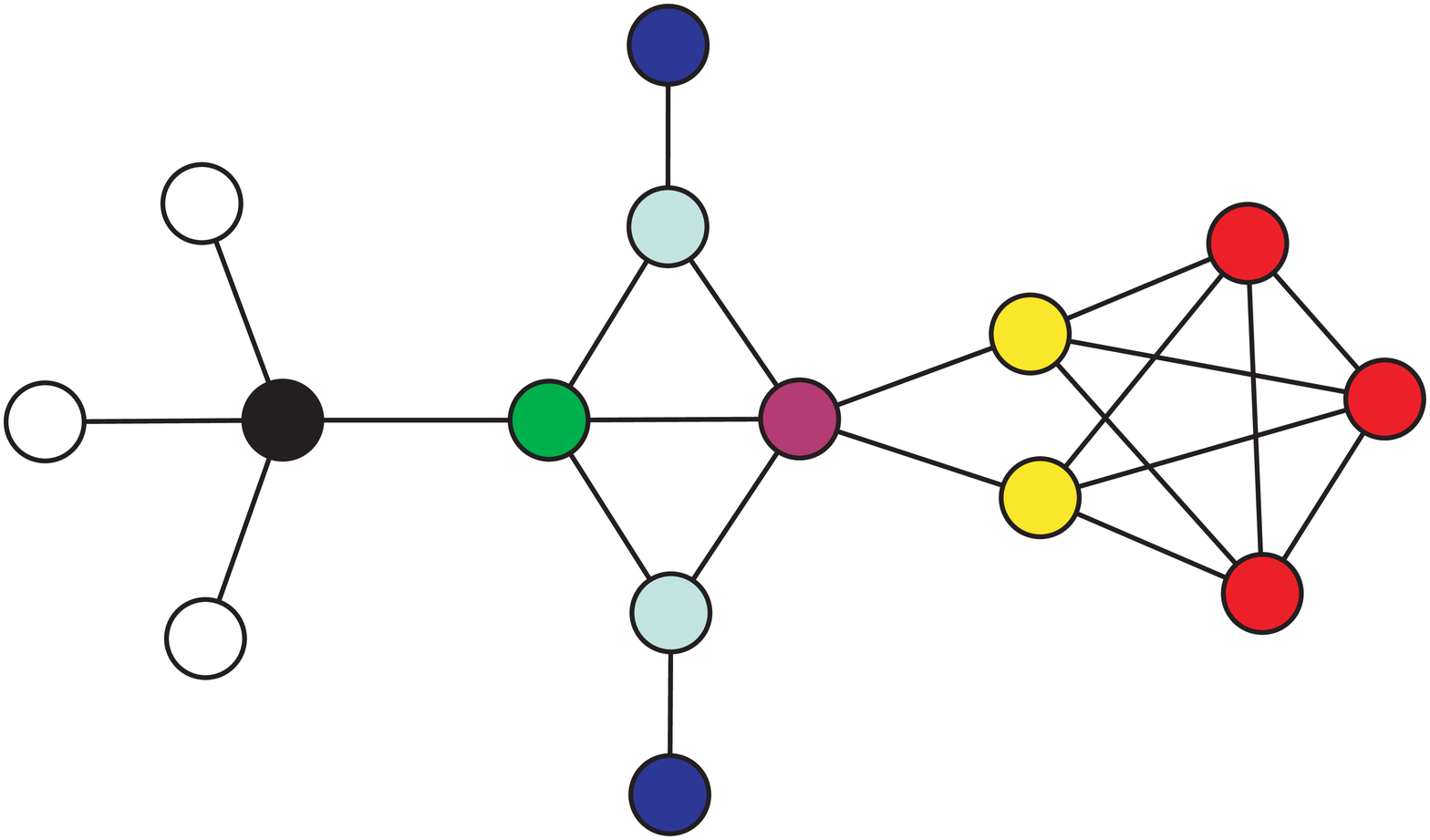}} 
 \subfigure[]{ \label{fig:quoo}
\includegraphics[scale=0.15]{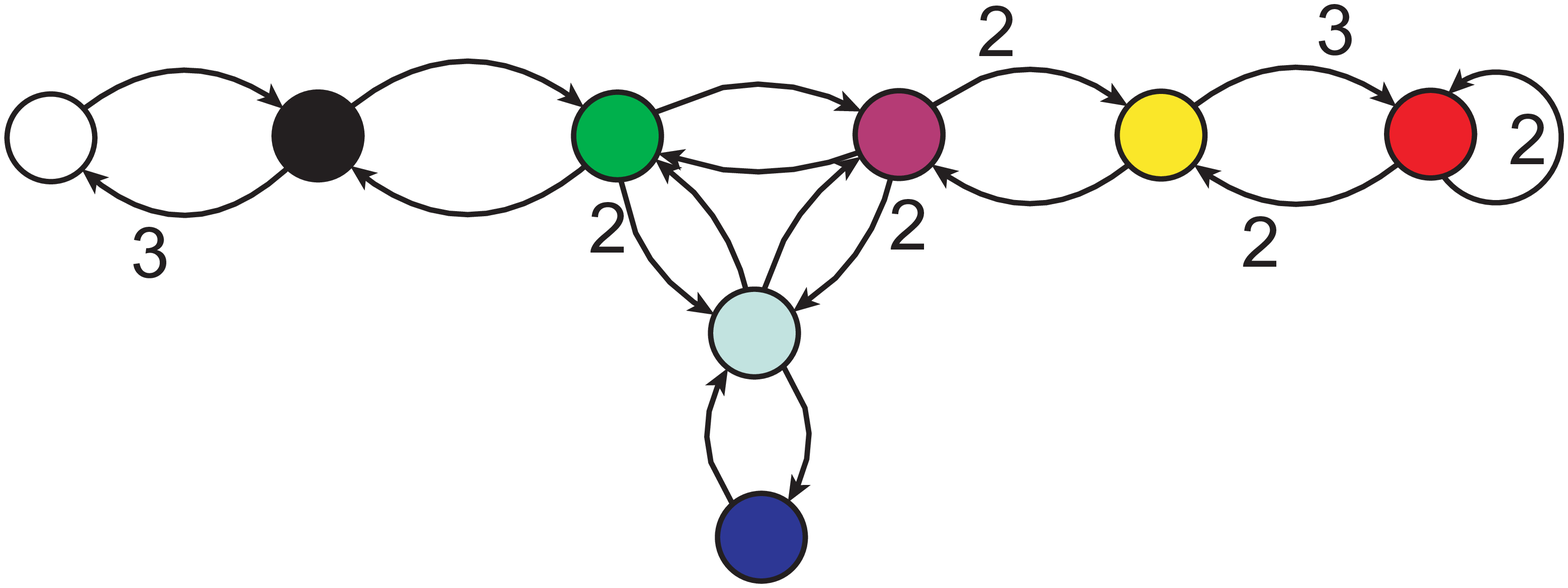}}
 \subfigure[]{ \label{fig:squo}
\includegraphics[scale=0.15]{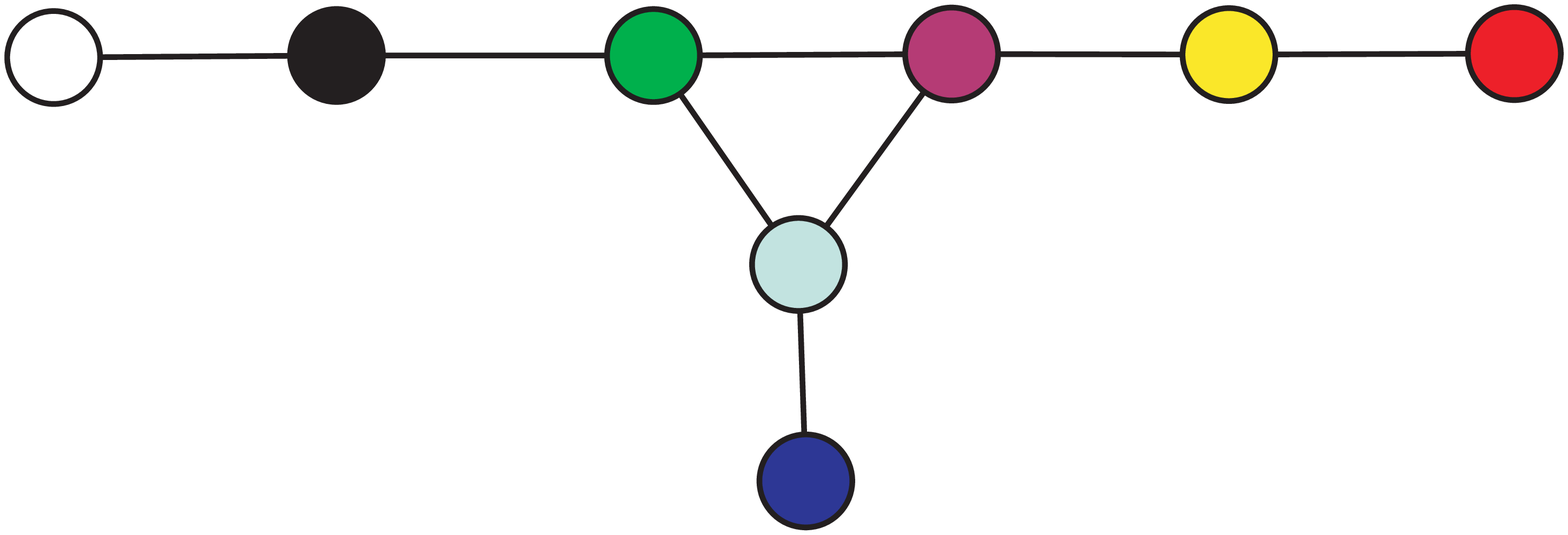}}
 \subfigure[]{ \label{fig:phd:parent}
\includegraphics[scale=0.3]{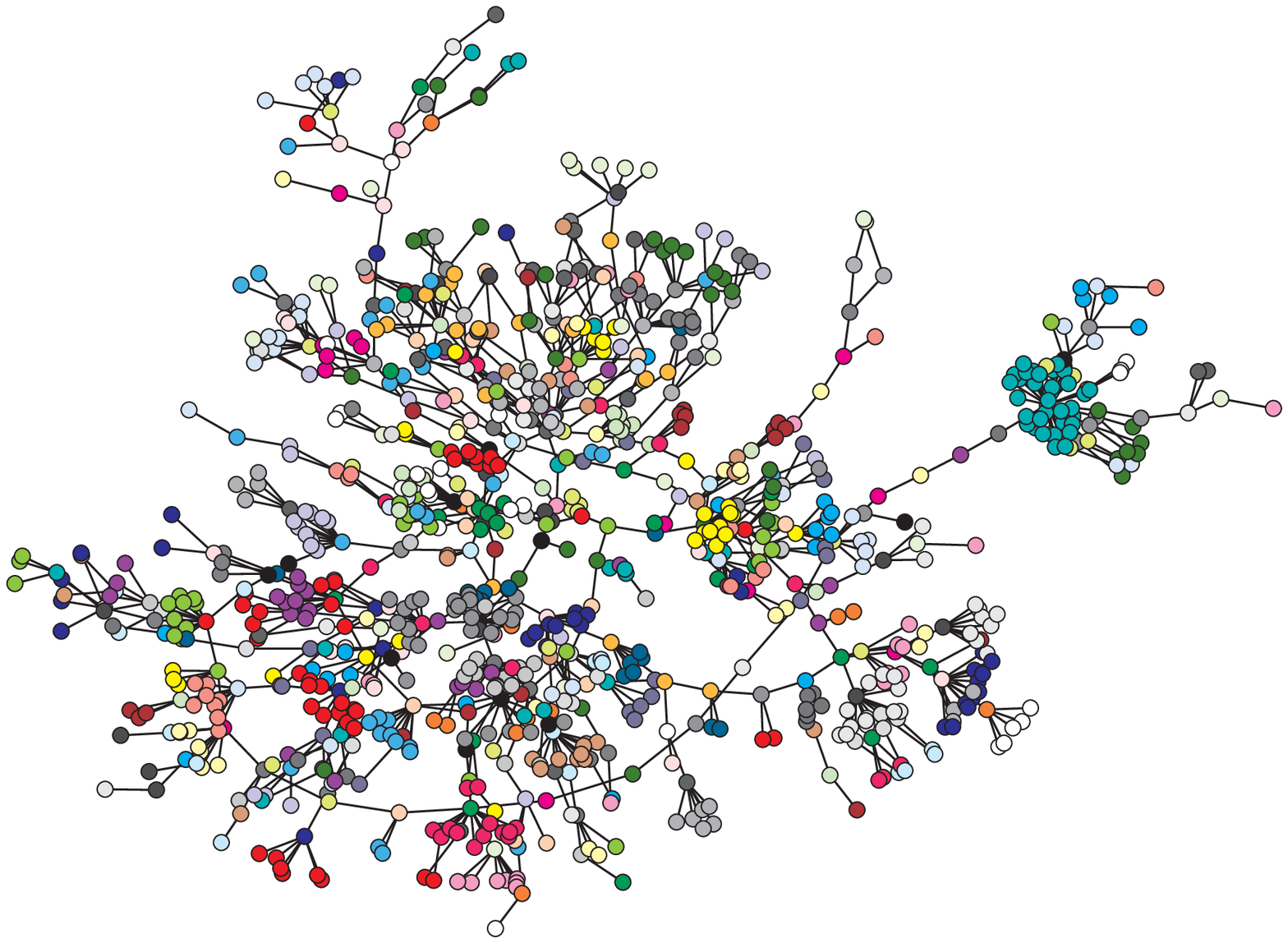}}
 \subfigure[]{ \label{fig:phd:quotient}
\includegraphics[scale=0.3]{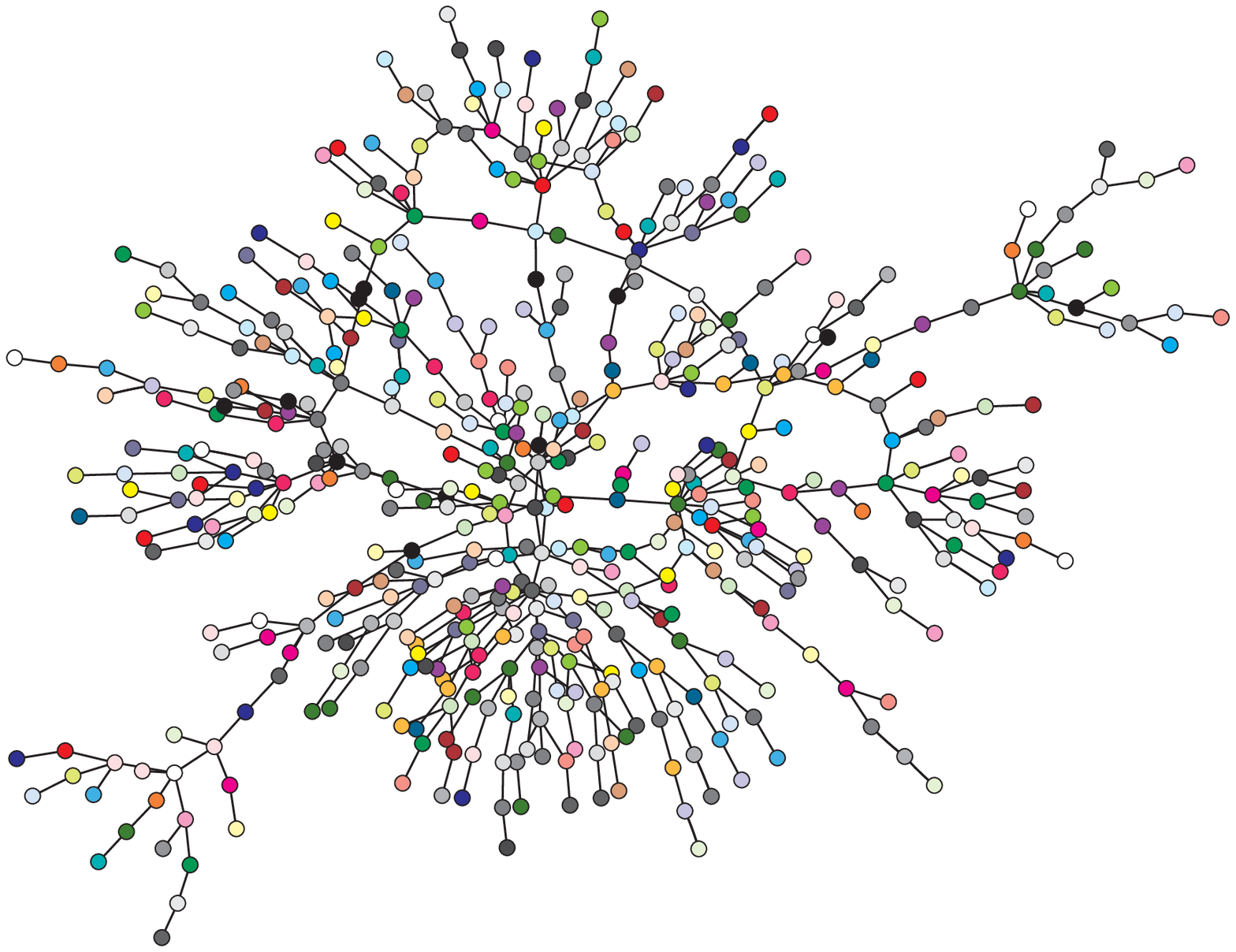}}

\caption{\small{\textbf{Networks and their quotients} (a) A
hypothetical network; (b) its quotient and (c) its s-quotient. (d) A
real-world example: the network of ties between Ph.D. students and
their advisers in theoretical computer science\cite{Johnson}. Each
edge links an adviser to a student. Vertices in the same orbit are
the same color. (e) The s-quotient of the theoretical computer
science network: vertices are colored as the orbits in the parent
network. There are 1025 vertices in the parent and only 511 in the
s-quotient (a reduction of 50.15\%); similarly there are 1043 edges
in the parent and only 525 in the s-quotient (a reduction of
49.67\%). Emprical networks were visualized using
Pajek\cite{pajek}.}} \label{fig:quo}
\end{figure*}

When a given network is a multi-digraph it is often convenient to
consider properties of the simple underlying network, in which edge
weights and directions are removed. Such underlying networks carry
the adjacency information of the full network, and so retain many
key network properties. Therefore as well as examining properties of
the quotient we shall also focus on properties of the simple
underlying quotient (or \emph{s-quotient} for short), denoted $Q_S$,
in which edge-directions, edge-weights and loops are removed from
the quotient. Fig. \ref{fig:quo} shows a network, its quotient and
its s-quotient. The s-quotient has the advantage that it retains the
adjacency information of the quotient, yet has a binary symmetric
adjacency matrix and thus is more computationally efficient to work
with.

\begin{table*}
\caption{\label{tab:skeleton}\textbf{Statistics for representative
networks and their s-quotients}. Summarized statistics for each
network are: the number of nodes $N$; the number of edges $M$; the
mean vertex degree $Z$; assortative mixing coefficient
$r$\cite{Newman2002}; the mean geodesic distance $m$; the diameter
$D$; and the clustering coefficient $C$\cite{watts}. The subscript
$s$ indicates quantities for s-quotients. We also calculate the
ratio of $N_s$ to $N$; the ratio of $M_s$ to $M$; and the ratio of
$M-M_s$ to $N-N_s$, denoted $z$. In all cases, we consider
properties of the underlying graph of largest connected component of
the parent networks. Except for PPI, InternetAS and Homo, all
network data can be downloaded from
\texttt{http://vlado.fmf.uni-lj.si/pub/networks/data/}. }
\begin{ruledtabular}
\begin{tabular}{cccccccccccccccccc}

Network&$N$&$N_s$&$N_s/N$&$M$&$M_s$&$M_s/M$&$z$&$Z$&$Z_s$&$r$&$r_s$&$m$&$m_s$&$D$&$D_s$&$C$&$C_s$\\
 \hline
California\cite{carlifornia}&5925&4009&67.66\%&15770&12882&81.69\%&1.51&5.32&6.43&- 0.23&-0.18&5.02&4.66&13&13&0.08&0.09\\
DutchElite\cite{dutchelite}&3621&1907&52.67\%&4310&2576&59.77\%&1.01&2.38&2.70&-0.24&-0.04&8.56&7.71&22&22&0.00&0.00\\
Epa\cite{epa}&4253&2212&52.01\%&8897&6545&73.56\%&1.15&4.18&5.92&-0.30&-0.16&4.50&4.11&10&10&0.07&0.10\\
Erdos02\cite{erdos}&6927&2365&34.14\%&11850&7034&59.36\%&1.06&3.42&5.95&- 0.12&-0.08&3.78&3.41&4&4&0.12&0.29\\
Eva\cite{eva}&4475&898&20.07\%&4652&1056&22.70\%&1.01&2.08&2.35&-0.19&0.00&7.53&7.43&18&18&0.01&0.05\\
Geom\cite{geom}&3621&2803&77.41\%&9461&7346&77.65\%&2.59&5.23&5.24&0.17&0.19&5.31&5.15&14&14&0.54&0.43\\
Homo\cite{biogrid}&7020&6066&86.41\%&19811&18575&93.76\%&1.30&5.64&6.12&- 0.06&-0.01&4.86&4.77&14&14&0.10&0.11\\
P-fei1738\cite{p-fei}&1738&1176&67.66\%&1876&1312&69.94\%&1.00&2.16&2.23&-0.27&0.07&10.22&10.39&29&29&0.00&0.00\\
PPI\cite{hjong}&1458&1019&69.89\%&1948&1469&75.41\%&1.09&2.67&2.88&-0.21&-0.05&6.80&6.68&19&19&0.07&0.07\\
Yeast\cite{yeast}&2284&1852&81.09\%&6646&6138&92.36\%&1.18&5.82&6.63&- 0.10&-0.04&4.15&4.17&11&11&0.13&0.14\\
InternetAs\cite{caida}&22442&11392&50.76\%&45550&29564&64.90\%&1.45&4.06&5.19&-0.20&-0.19&3.86&3.86&10&10&0.22&0.20\\
\end{tabular}
\end{ruledtabular}
\end{table*}

\section{Properties of Quotients}

\subsection{Relative size}
Since quotients and s-quotients are formed by factoring out network
redundancy they can be significantly smaller than their parent
networks. Table \ref{tab:skeleton} shows that many empirical
s-quotients are less than 50\% the size of their parent network,
illustrating that much real-world network structure is due to
repetition of structurally identical elements.

In order to investigate relative sizes of s-quotients we examined
the correlation between various measures of network symmetry and the
ratio of the size of the s-quotient to that of its parent (the
reduction ratio). We used two different indicies to quantify network symmetry:
(1) $\beta_G$, the $n$th root of the ratio of automorphism group
size to that of the (maximally symmetric) complete graph of the same
size\cite{sym1,sym2}:
\[
\beta_G=\bigg ( \frac{|\textrm{Aut}(G)|}{N!} \bigg )^{1/N}
\]
and (2) $\gamma_G$, the ratio of the number of vertices in
non-trivial orbits to $N$, the number of vertices in the
network\cite{xiao1,xiao2}:
\[
\gamma_G=\frac{\sum_{|\Delta_i|>1}|\Delta_i|}{N}.
\]
We define the size of a network as $|G|=N+M$, where $M$ is the
number of edges in $G$. The quotient reduction ratio is defined as
$r_G=|Q_S|/|G|$.

\begin{figure}
\centering
\includegraphics[scale=0.6]{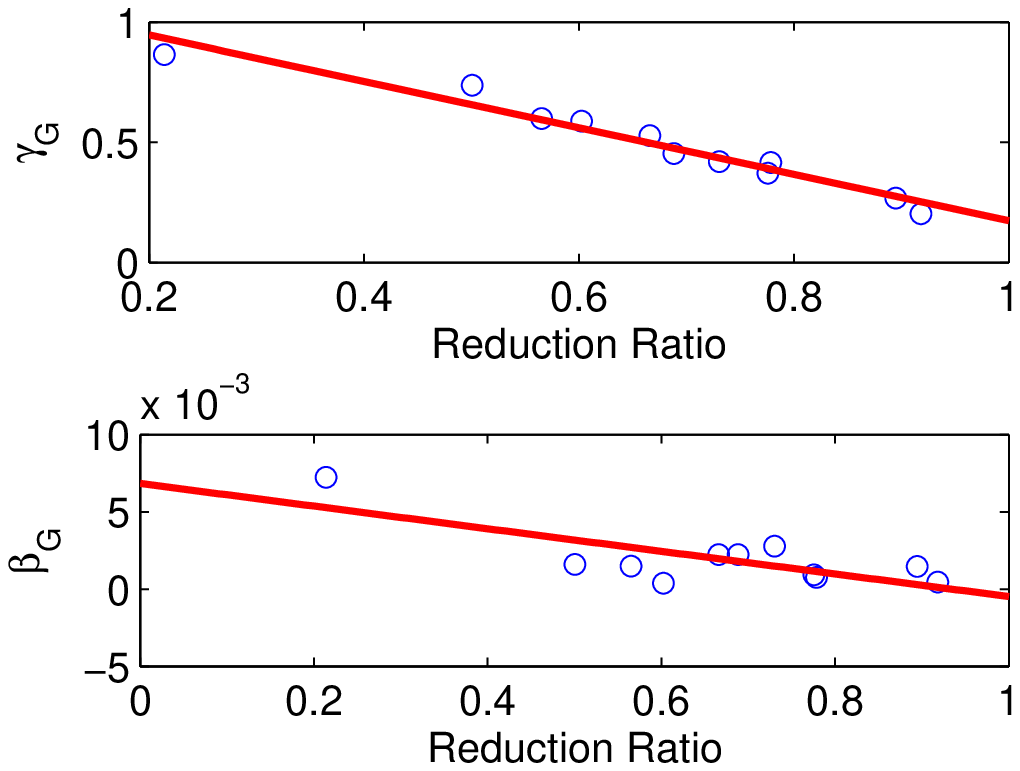}
\caption{\textbf{Network symmetry and s-quotient size are inversely correlated}.}.
\label{fig:corre_red_symmetry}
\end{figure}

Fig. \ref{fig:corre_red_symmetry} shows the correlation between
these two measures of symmetry and the quotient reduction ratio for
eleven representative real-world networks (further details of these
networks are given in Table \ref{tab:skeleton}). The correlation
coefficient between $r_G$ and $\beta_G$ is -0.7567; the correlation
coefficient between $r_G$ and $\gamma_G$ is -0.9767, illustrating
that the degree of symmetry and the relative size of the s-quotient
are strongly negatively correlated over a variety of networks.

\subsection{Heterogeneity}
Network heterogeneity -- that is, the degree to which different
vertices play different roles or possess different properties in a
network -- is important in determining many dynamic
network properties such as robustness\cite{Albert} and
synchronization\cite{nish}. A completely heterogeneous network is
one in which all vertices play a unique structural role (that is,
the network has a trivial automorphism group), while a completely
homogeneous network is one in which all vertices play the same
structural role (that is, the network has a transitive automorphism
group)\cite{xiao1}. Since structurally equivalent elements are
removed in the quotient while structurally non-equivalent elements
are preserved it is immediate that network quotients are completely
heterogeneous: all vertices in the quotient play a different
structural role (see the quotient in Fig. \ref{fig:quo} for
example). However, since edge weights, directions and loops are
removed in the s-quotient, some vertices may still play the same
structural role (for example in the s-quotient in Fig. \ref{fig:quo}
the red and white vertices are structurally equivalent; as are the
yellow and black vertices; as are the green and purple vertices).
Thus, although s-quotients may not be completely heterogeneous we
expect that they will be more heterogeneous than their parent
networks.

\begin{figure}
\centering
\includegraphics[scale=0.5]{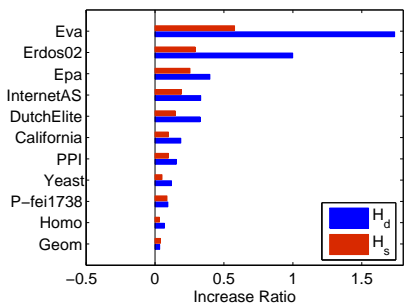}
\caption{\textbf{Heterogeneity of empirical networks and their
s-quotients}. The horizontal axis shows the ratio of heterogeneity
in the s-quotient to that of its parent, defined as
$\bar{H}_{d,s}(Q_S)/\bar{H}_{d,s}(G)-1$}. \label{fig:hetero}
\end{figure}

In order to assess network heterogeneity we used two distinct
measures: degree-based entropy\cite{wangb} $H_d(G)$ and
symmetry-based entropy\cite{xiao1} $H_s(G)$. These two entropies
have a common algebraic form:
\[
H_{d,s}(G)=-\sum_{i} {p_i}\log{p_i},
\]
where $p_i$ is the probability that a vertex has
degree $i$ when calculating $H_d(G)$; and $p_i$ is the probability
that $v \in \Delta_i$ when calculating $H_s(G)$. In order to compare
networks of different sizes we normalized these measures as follows:
\[
\bar{H}_{d,s}(G)=\frac{H_{d,s}(G)-\min(H_{d,s},N)}{\max(H_{d,s},N)-\min(H_{d,s},N)}.
\]
where $\max(H_{d,s},N)$ and $\min(H_{d,s},N)$ are the maximal and
minimal entropy values for a network with $N$ vertices respectively.
Fig. \ref{fig:hetero} summarizes these two entropy measures for the
$11$ empirical networks in Table \ref{tab:skeleton}. As expected in
all cases the s-quotient is more heterogeneous than its parent
indicating that structural features which contribute to network
homogeneity are factored out in the s-quotient while structural
features which contribute to network heterogeneity are preserved in
the s-quotient.

\subsection{Vertex degree distributions}
A quotient's vertex degree distribution is strongly related to that
of its parent. Recall that all vertices in the same orbit have the
same degree\cite{lauri}. Thus, $n_{\textrm{out}}(k,Q) =  O_k $,
where $O_k$ is the number of orbits of degree $k$ in $G$ and
$n_{\textrm{out}}(k,Q)$ is the number of vertices in $Q$ with
out-degree $k$. We may think of the vertex out-degree distribution
in the quotient as being formed by measuring the degree of one
representative vertex from each orbit in $G$. Thus, the quotient
vertex out-degree distribution represents the `essential' vertex
degree distribution of its parent and is dependent upon both the
vertex degree distribution and the symmetry structure of its parent.

Hub vertices (those with high degree) often dominate real-world
network topology, and consequently crucially affect network
properties such as robustness\cite{Albert} and traffic along
geodesics\cite{goh}. Hence, in order to accurately preserve network
properties, quotients should preserve hub vertices. Since they
generally connect many disparate regions of a network, hub vertices
are more likely to be fixed by the automorphism group than are
vertices of low degree, and consequently we expect that generically
this is indeed the case.

Much of the symmetry present in many real world networks is due to
the presence of bicliques\cite{sym2, xiao2} and in particular, the
presence of stars\cite{sym1,sym2} (a $k$-star is a subgraph
consisting of a central vertex of degree $>k$ adjacent to $k$
vertices of degree $1$). In $k$-stars, the $k$ vertices of degree 1
are structurally equivalent to each other and collapse to a single
vertex in the quotient. Thus, each $k$-star reduces the vertex order
of the quotient by $k-1$. Fig \ref{fig:quo} shows how a 3-star (in
white on the left) collapses to a single vertex in the quotient and
s-quotient. In networks which contain a significant number of
bicliques or stars, the s-quotient is formed largely by `pruning'
appropriate vertices of small degree from the parent network while
fixing hubs.

\begin{figure}
\centering
\includegraphics[scale=0.5]{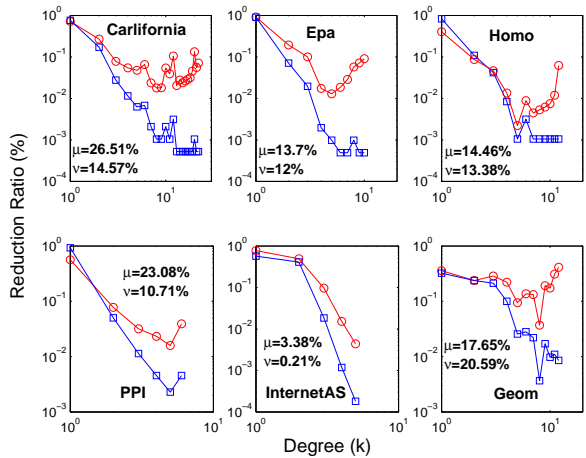}
\caption{\textbf{Factored degree distributions in s-quotients}. The
symbol $\bigcirc$ gives the reduction ratio $P_k$; the symbol $\Box$
gives the reduction ratio $R_k$.} \label{fig:reduction}
\end{figure}

In order to assess the degree to which hub vertices are preserved in
quotients, and the degree to which quotients are formed by pruning
vertices of low degree, we investigated the degree distributions
of those vertices that have been factored out in s-quotients for a
variety of real-world networks. In particular, we considered two
distinct quantities: (1) $P_k$ the number of vertices of degree $k$
factored out in the s-quotient as a percentage of the total number
of vertices in the parent network:
\[
P_k = \frac{N_k-O_k}{N_k} \times 100 \%,
\]
where $N_k$ is the number of vertices with degree $k$ and $O_k$ is
the number of orbits in which each vertex has degree $k$; and (2)
$R_k$ the number of vertices of degree $k$ factored out in the
s-quotient as a percentage of the total number of vertices factored
out:
\[
R_k = \frac{N_k-O_k}{N-|\mathbf{\Delta}|} \times 100 \%.
\]
Fig. \ref{fig:reduction} shows that generally in empirical networks
only vertices with small degree are factored out in the s-quotient
(in all tested cases the maximum degree of any factored vertex was
29); vertices with higher degree tended to be fixed by the
automorphism group and thus retained in the s-quotient.

In order to identify the proportion of vertices which are factored
out by degree we considered two further quantities. Let $d(v)$ be
the degree of a vertex $v$. Consider now the total network
degree-set: $\textrm{Deg}=\{d(v)|v\in G\}$ (i.e. the set of all
vertex degrees), and the nontrivial orbit degree-set:
$\textrm{Deg}'=\{d(v)|v\in \Delta_i, |\Delta_i|>1 \}$ (i.e the set
of degrees of those vertices in nontrivial orbits). Note that
$\textrm{Deg}' \subseteq \textrm{Deg}$. We define two quantities
based upon these sets:
\[
\mu=\frac{|\textrm{Deg}'|}{|\textrm{Deg}|} \times 100 \%,
\]
the percentage of the degree-set factored out in the s-quotient, and
\[
\nu=\frac{\max(\textrm{Deg}')}{\max(\textrm{Deg})} \times 100\%,
\]
the maximum vertex degree factored out in the s-quotient as a
percentage of the maximum vertex degree in the parent network. Fig.
\ref{fig:reduction} shows these measures for $6$ real-world
networks. It is clear that vertices which are factored out in the
s-quotient constitute only a minority of the whole network-degree
set (the maximum value for $\mu$ we found was $26.51\%$); and that
only vertices with relatively low degree are factored out (the
maximum value for $\nu$ we found was $20.59\%$). Furthermore, Table
\ref{tab:skeleton} also shows that it is common for s-quotients to
have an average degree larger than that of their parent,
demonstrating that vertices of small degree are more likely to lie
in a non-trivial orbits (and thus be factored out in the quotient)
than are hub vertices (which are generically retained).

\subsection{Communication properties}
Many empirical large complex networks are `small-world' meaning that
there exists a relatively short path between any two vertices in the
network\cite{newman}. The shortest path between a pair of vertices
is known as a geodesic and the length of the longest geodesic is
known as the diameter of the network, which we denote $D(G)$.
Distribution of geodesic distances and network diameter both
significantly effect dynamic network properties such as information
transfer\cite{tree} and tolerance to attack\cite{Albert}.

Table \ref{tab:skeleton} shows the network diameter and s-quotient
diameter for a variety of empirical networks. In all cases, network
diameter is preserved exactly in the s-quotient. For example, in the
Eva network, a telecommunications and media ownership
network\cite{eva}, the vertex and edge numbers of the s-quotient are
$20\%$ and $22.7\%$ that of original network respectively, yet
network diameter is maintained in the s-quotient. In this case the
s-quotient is substantially smaller than its parent, yet it
preserves the communication properties of its parent. In fact, this
empirical observation is true for all `locally-symmetric' networks.

Intuitively, a network is \emph{globally-symmetric} if the longest
geodesic is between vertices in the same orbit (that is, there are
automorphisms which permute distant vertices); otherwise the network
is \emph{locally symmetric} (that is, all automorphisms act on local
vertex subsets). Since many real-world networks are commonly subject to
continuous stochastic fluctuations in topology, we do not expect --
neither did we find -- that any large real-world networks are
globally symmetric.

The s-quotient describes the orbit adjacency structure of its parent
network. Thus, network diameter is exactly preserved in the
s-quotient as long as the parent network is not globally symmetric.
For an illustration of this see the network shown in Fig.
\ref{fig:net}. In this network the longest geodesic is between any
of the red vertices on the right and any of the white vertices on
the left, and the network has diameter = 5. The s-quotient of this
network (shown in Fig. \ref{fig:squo}) also has diameter = 5, and
diameter is preserved in the s-quotient since orbit adjacency is
preserved.

\begin{figure}
\centering
\includegraphics[scale=0.5]{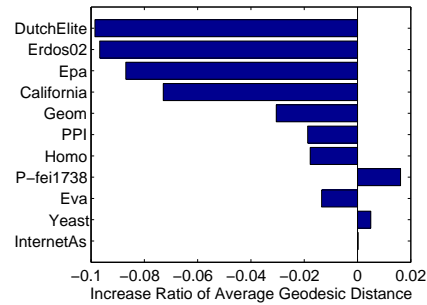}
\caption{\textbf{Mean geodesic distances in s-quotients}. The ratio
$(m_s/m)-1$ is plotted, where $m_s$ and $m$ are the mean geodesic
distances in the s-quotient and its parent respectively.}.
\label{fig:agd}
\end{figure}

 While the diameter of a network is related to the maximum
information transfer cost in the network, mean geodesic distance is
related to the average transit cost. Empirical measurements show
that the disparity between mean geodesic distance in the s-quotient
and its parent is usually quite small. As shown in Figure
\ref{fig:agd} for all tested networks the mean geodesic distance of
the s-quotient is within $10\%$ of that of its parent network
irrespective of the relative size of the s-quotient to its parent.
Since both network diameter and mean geodesic distance are robustly
inherited by the s-quotient from its parent, we conclude that the
s-quotient forms the communication skeleton of its parent.

\section{Conclusions}
The quotient of a network is formed by associating structurally
equivalent vertices into disjoint equivalence classes and
considering adjacency relationships between these equivalence
classes. Thus quotients capture all essential network complexity,
yet formally exclude all structural redundancy. Quotients may
therefore be thought of as the structural skeletons of the systems
from which they are derived. Consequently properties of the
quotient, and not those of the parent network \emph{per se},
describe core system complexity. Observation of the statistics of
real-world networks verifies that elements which contribute to
network homogeneity (or simplicity) are removed in network
quotients; while the elements which contribute to the heterogeneity
(or complexity) are completely retained in quotients.

Many biological networks are thought to form by growth with vertex
duplication, or partial duplication\cite{chung}. Vertex duplication
is useful in a biological context since it naturally endows
biological regulatory systems with functional redundancy, thus
reinforcing against damage. Quotients of biological regulatory
networks therefore encode core relationships between biochemical
control motifs, minus any repetitions due to redundancy. While the
large-scale properties of analogous biological regulatory networks
are often remarkably similar across a broad range of species, their
detailed properties can differ significantly\cite{biogrid,jeong}.
Thus it may be of particular interest to explore the similarities
between structural properties of quotients of regulatory networks
for various different species, since this could provide a new means
to analyze functional conservation of regulatory motifs across
species.

 Finally, since quotients carry the
structure of their parents, yet are often substantially smaller,
performing analysis directly on quotients, rather than on the
corresponding parent networks, can reduce the complexity of network
algorithms. For example, average shortest path length computation
time can be reduced from $\Theta(NM)$ to $\Omega(r_Nr_MNM)$ if
calculated on the s-quotient (where $r_N=N_s/N$ and $r_M=M_s/M$).

We anticipate that further investigation of properties of network
quotients will be both of theoretical and pragmatic interest.

\section{\label{sec:level1}Acknowledgments}
The work was supported by the National Natural Science Foundation of
China under Grant No.60303008, No.60673133, No.60703093; the
National Grand Fundamental Research 973 Program of China under Grant
No.2005CB321905.

\section{\label{sec:level2}References and notes}

\end{document}